\documentclass[runningheads]{llncs}
\usepackage{graphicx}
\usepackage{textcomp}
\usepackage{booktabs,array} 
\usepackage{tablefootnote}
\begin{document}
\title{Coronary Artery Plaque Characterization from CCTA Scans using Deep Learning and Radiomics}
\titlerunning{Plaque Characterization with DL and Radiomics}
%
\author{Felix Denzinger\inst{1,2}
	\and
Michael Wels\inst{2} \and 
Nishant Ravikumar\inst{1} \and
Katharina Breininger\inst{1} \and
Anika Reidelsh\"ofer\inst{3} \and
Joachim Eckert\inst{4} \and
Michael S\"uhling \inst{2} \and
Axel Schmermund \inst{4} \and
Andreas Maier \inst{1}
}
\authorrunning{ Felix Denzinger et al.}
\institute{Pattern Recognition Lab, Friedrich-Alexander-Universität Erlangen-Nürnberg, Erlangen, Germany \and
Computed Tomography, Siemens Healthcare GmbH, Forchheim, Germany\and
University Clinic Frankfurt, Frankfurt am Main, Germany\and
Cardioangiological Centrum Bethanien, Frankfurt am Main, Germany
}
%
\maketitle              
%

\hyphenation{angio-graphy}
\hyphenation{tomo-graphy}

\begin{abstract}
Assessing coronary artery plaque segments in coronary CT angiography scans is an important task to improve patient management and clinical outcomes, as it can help to decide whether invasive investigation and treatment are necessary.
In this work, we present three machine learning approaches capable of performing this task. 
The first approach is based on radiomics, where a plaque segmentation is used to calculate various shape-, intensity- and texture-based features under different image transformations. 
A second approach is based on deep learning and relies on centerline extraction as sole prerequisite.
In the third approach, we fuse the deep learning approach with radiomic features.
On our data the methods reached similar scores as simulated fractional flow reserve (FFR) measurements, which - in contrast to our methods - requires an exact segmentation of the whole coronary tree and often time-consuming manual interaction. 
In literature, the performance of simulated FFR reaches an AUC between 0.79-0.93 predicting an abnormal invasive FFR that demands revascularization.
The radiomics approach achieves an AUC of 0.86, the deep learning approach 0.84 and the combined method 0.88 for predicting the revascularization decision directly.
While all three proposed methods can be determined within seconds, the FFR simulation typically takes several minutes.
Provided representative training data in sufficient quantities, we believe that the presented methods can be used to create systems for fully automatic non-invasive risk assessment for a variety of adverse cardiac events.

\end{abstract}
%
%
%
\section{Introduction}
Cardiovascular diseases (CVDs) have persisted to be the leading cause of death across all developed countries~\cite{Mendis15}. 
Most CVDs are related to atherosclerotic plaques in the associated arteries~\cite{naghavi03}.
Two types of high risk plaque segments exist: functionally significant plaques, which narrow the lumen and immediately lead to cardiac ischemia, and vulnerable plaques, which can rupture and cause thrombus formation and adverse coronary syndromes (ACS) such as stroke or cardiac infarction.\\
The reference standard measure to judge whether a plaque segment is functionally significant and the corresponding vessel needs to be revascularized is the fractional flow reserve (FFR) value. FFR is defined as the pressure after a lesion relative to the pressure before the lesion, and is measured invasively~\cite{Cury16}.
As interventional procedures involving the heart have the risk of inducing adverse cardiac events, a non-invasive assessment of the type of plaque for further patient selection is highly desirable. 
A non-invasive approach for this task is simulated FFR, which aims to simulate the FFR values from coronary computed tomography angiography (CCTA) data using a fluid dynamics approach~\cite{taylor13}, which requires an exact segmentation of the whole coronary tree and computational mesh generation~\cite{wels16}. Sufficient segmentation quality can often only be achieved with time-consuming manual interaction.\\
Previously, radiomics have been proposed to represent quantitative image information which is inherent in the data but hard to interpret for human readers~\cite{lambin12}. 
They include multiple intensity-, texture-, shape- and transformation-based metrics extracted from a lesion segmentation and have been shown to be able to characterize coronary plaques~\cite{kolossvary17}. 
More recently, deep learning has been investigated to detect lesions with a high stenosis degree and to categorize the calcification grade of coronary plaques using a recurrent convolutional neural network (RCNN)~\cite{zreik18b}. 
In their work, they first extract multi planar reformatted (MPR) slices orthogonally to each centerline point.
Next, they cut the resulting image stack into multiple overlapping cubes from which features are extracted using a 3D convolutional neural network (CNN). 
Finally, classification is achieved using a sequence analysis network. \\
In this work, we propose a fully automatic method to directly predict the clinical decision of revascularization based on single plague segments. 
We investigate three machine-learning approaches for classification: radiomic feature analysis, deep learning and a combination of both. 
For the first variant, radiomic features are extracted from each vessel segment based on the vessel segmentation in the region of interest. 
Contrary to the approach in~\cite{kolossvary17}, we do not perform data mining since it neglects cross-feature correlations. Instead, we train a bagging classifier, namely the XGBoost algorithm~\cite{chen16}, which automatically detects relevant features and uses all information from the features. 
For the deep learning approach, we extend the approach presented in~\cite{zreik18b} by improving the data representation using a transformation of the image stack into a cylindrical coordinate system which allows for a more effective training of the network and reduce the risk of overfitting by using 2D instead of 3D convolutions. 
Thirdly, we propose a novel combination of both aforementioned approaches. 
After extracting a sequence of cubes along the centerline, we calculate the radiomic features of each cube using a plaque segmentation mask extracted a priori. The resulting sequence of radiomic features is then recombined with a multi-layer perceptron (MLP) and subsequently analyzed using a sequence analysis network based on gated recurrent units (GRUs).
We evaluate all variants on CCTA scans of 95 patients with a total of 345 plague segments using ten-fold cross validation and compare our results with simulated FFR.

\section{Data}
The data collection contains CCTA scans of 95 patients taken within a time span of 2 years with the same system. 
The decision for revascularization or further invasive assessment was based on different clinical indications, e.g., functional tests including cardiac stress MRI or MIBI SPECT, and was made by trained cardiologists. 
In some cases, identification of culprit lesions was additionally based on ECG abnormalities if these indicated a bad perfusion of a specific part of the heart muscle.
In total, the data contained 345 lesions, which were annotated by defining their start and end centerline point and segmenting their inner and outer vessel wall using a fully automatic approach~\cite{lugauer14}.
For all data sets, automatic centerline extraction was performed as described in~\cite{zheng13}.
For each main branch of the coronaries a label indicated whether it was revascularized or not. 
To obtain reliable labels on the segment level, we propagated a positive revascularization decision only to the segment with the highest stenosis grade. In order to allow for an comparison with the results in~\cite{zreik18b}, we additionally assessed for all segments whether the stenosis grade was below or above 50$\,\%$. With this procedure 85 (24.64\,\%) lesions were labeled as having a high stenosis grade and 93 (26.97\,\%) as requiring revascularization.

\section{Methods}

\subsection{Radiomic-based Classification} 
As mentioned, a multitude of shape-, intensity- and texture-based features is extracted under different image transformations from the lesion segmentation as radiomics. 
A detailed description of all radiomic features can be found in~\cite{kolossvary18}.
The extracted feature vector has a high dimensionality. Therefore, direct classification is hard to achieve due to the curse of dimensionality. 
To overcome this we used the XGBoost classifier~\cite{chen16}, which calculates its prediction based on an ensemble of decision trees while minimizing a loss function based on the total ensemble prediction.
Since new leaves are added based on greedy search, only relevant non-redundant features are selected during training.
Features were calculated using the open source PyRadiomics library~\cite{griethuysen17} selecting all possible features under all transformations. 
\begin{figure}
\centering
\includegraphics[width=0.7\textwidth]{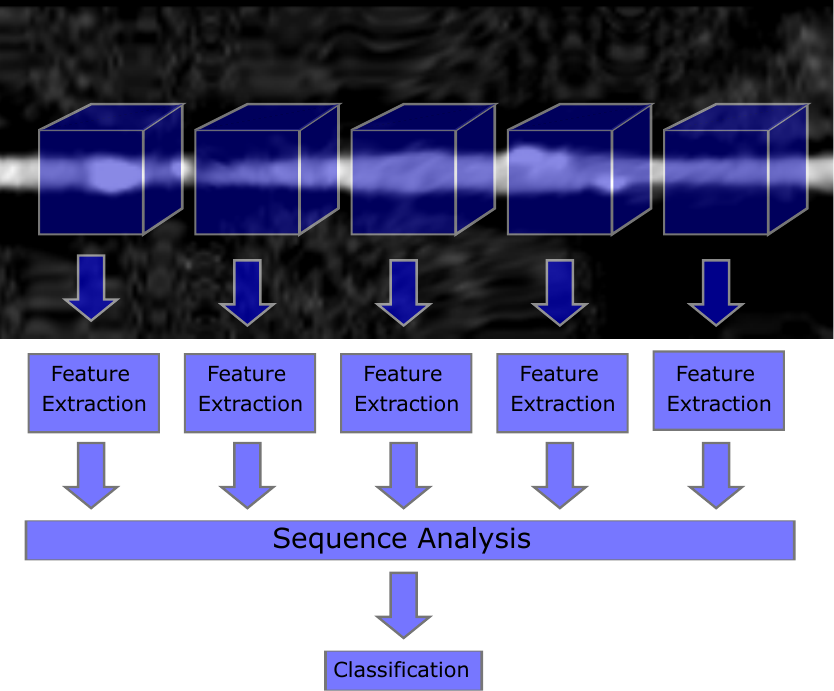}
\caption{Algorithm overview: extraction of a sequence of cubes along the centerline is followed up by a feature extraction method -- either with a convolutional neural network or the PyRadiomics module. 
	The resulting sequence of features is then analyzed by a sequence analysis neural network.} \label{fig1}
\end{figure}
\subsection{Deep Learning-based Classification}
\begin{figure}
\centering
\includegraphics[width=0.7\textwidth]{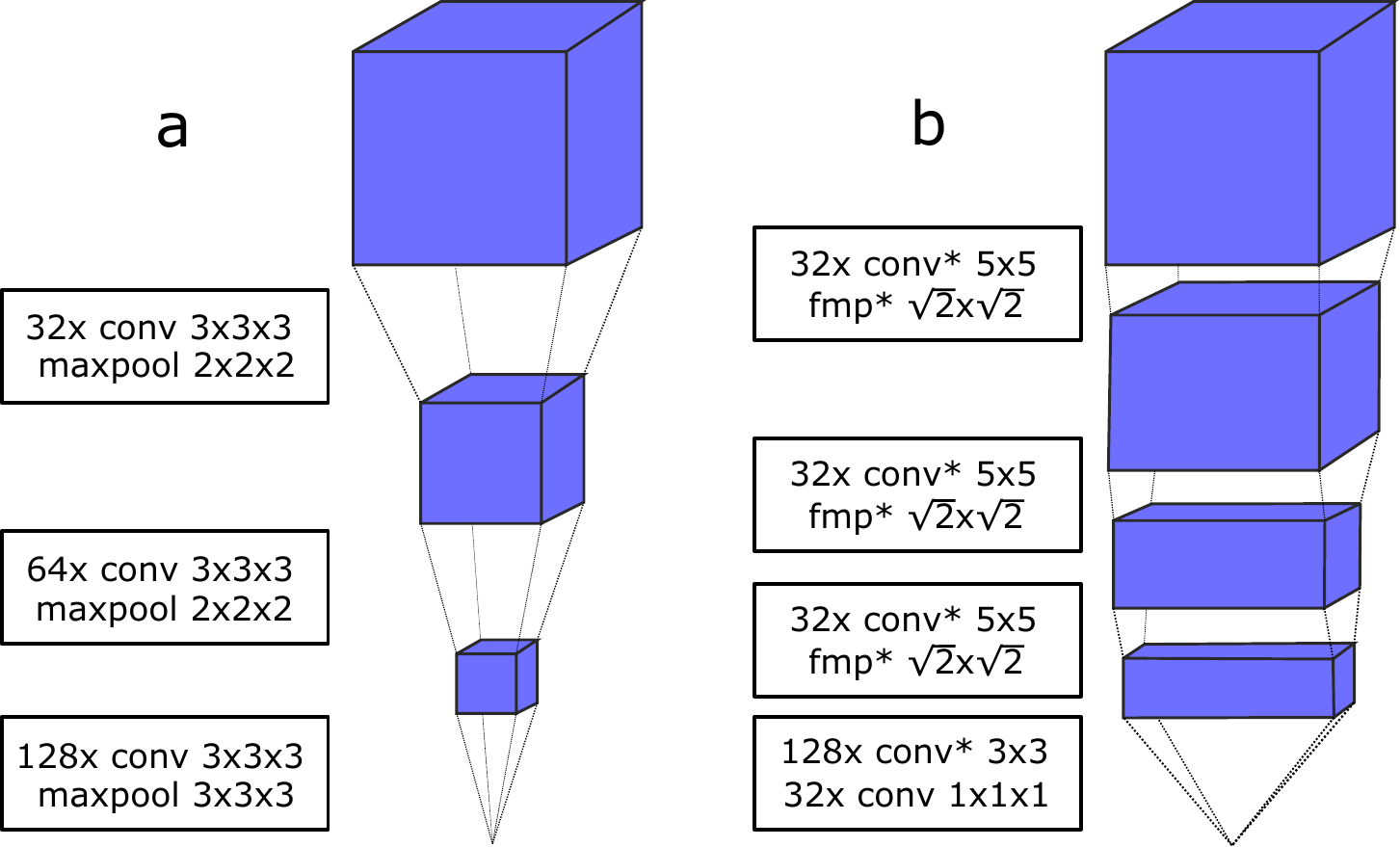}
\caption{Feature extraction model overview: a) model as described in~\cite{zreik18b}. b) our proposed model. * denotes a slice-wise operation, and fmp denotes fractional max pooling~\cite{graham14}, which allows a pooling size smaller than 2 which enables feature extraction from intermediate scales.} \label{fig2}
\end{figure}

The second approach is based on deep learning and can be separated into several steps: data extraction, local feature extraction and sequence analysis. An overview of the workflow is shown in Figure~\ref{fig1}.
First, MPR slices are extracted orthogonally to each point of the centerline in the segment. 
Then, the resulting image stack is cut into multiple overlapping cubes. 
The extracted cubes are transformed to polar coordinates to allow for a better representation for the neural network. 
The motivation behind this lies in the assumption that features that characterize lesions are formed radially to the centerline and vary along the centerline direction.
The slices of each transformed cube are then used as input to a 2D-CNN that performs a slice-wise feature extraction. This is followed by 1x1 convolutions in centerline direction that recombine and fuse the information across a cube to perform a local feature extraction. 
The architecture of the feature extraction network is depicted in Figure~\ref{fig2}, alongside the 3D-CNN network proposed in~\cite{zreik18b} that we evaluate for comparison.
To obtain a final classification, we perform a sequence analysis using a two layer recurrent neural network (RNN) using gated recurrent units~\cite{cho14} on the features extracted from the cubes with the centerline direction as ``temporal'' dimension. 
Based on the assumption that information about the plaque composition is contained in both directions of the centerline, we perform the sequence analysis in a bidirectional fashion. 

\subsection{Combined Approach} 
A common way to train neural networks with a limited amount of data is to use pretrained models, which comprise relevant image features already learned on different data sets.
However, this is difficult when dealing with medical data, since data of different organs, modalities and use cases are often not correlated and three-dimensional.
To overcome this, non-deep learning feature extraction methods can be used and combined with deep learning.
Therefore, we combine the above mentioned radiomic and deep learning approaches. 
Again, the vessel was sliced in a sequence of overlapping volume cubes, but now the feature extraction was performed using the PyRadiomics library and the vessel segmentation of the plaque segment. 
Since preliminary experiments suggested the shape-based feature group to be the most important for estimating both the revascularization decision and the stenosis degree, we focused on these features.
The resulting sequence of radiomic feature vectors was further evaluated using a three layer MLP before analyzing the sequence with bidirectional GRUs.

\section{Experiments and Results}
We evaluate the proposed approach for binary stenosis grade classification (high-degree stenosis $>$ 50$\,\%$, low-degree stenosis $<$ 50$\,\%$) to allow for a direct comparison with~\cite{zreik18b} and for the prediction of clinical revascularization decisions.
For all experiments, evaluations were performed using ten-fold cross validation with patient-wise stratified splitting. For the neural network based methods, 20$\,\%$ of the training data was set aside as validation set in each fold. For each fold, the networks were trained for 50 epochs and the model that performed best on the validation set was selected for evaluation on the test set. For the CNN-based methods, data augmentation was performed in form of random rotation, mirroring along the x-axis, translation and additive Gaussian noise, and we resampled the data during batch generation to achieve class balance.
In order to normalize our data, histogram equalization was performed for each approach before feature extraction.
To evaluate our approaches, we computed the area under the receiver operating characteristic curve (AUC), accuracy (Acc), F1-score, positive predictive value (PPV), negative predictive value (NPV), sensitivity, specificity and the Matthews correlation coefficient (MCC).

\subsection{Stenosis Grade Classification}
 
\begin{table}[t]
	\centering
	\caption{Evaluation results for stenosis degree prediction on lesion-level. The results in the first row are copied from~\cite{zreik18b}.}\label{stenosisResults}
	\begin{tabular}{ l | l | l | l | l | l | l | l | l }
		\hline
		Model/metric &  AUC & Acc & F1 & PPV & NPV & Sens & Spec & MCC\\
		\hline
		3D-RCNN \cite{zreik18b} (orig data) & --  & 0.94 & 0.64 & 0.65 & 0.97 & 0.63 & 0.97 & 0.60 \\
		\hline
		\hline
		3D-RCNN \cite{zreik18b} (our data) & 0.89 & 0.85 & 0.67 & 0.58 & \textbf{0.94} & 0.79 & 0.86 & 0.59 \\
		\hline
		2D-RCNN + polar transform& 0.86 & 0.87 & 0.64 & 0.68 & 0.91 & 0.60 & \textbf{0.93} & 0.56\\
		\hline
		Radiomics + XGBoost  & 0.89 & 0.84 & 0.69 & 0.69 & 0.90 & 0.68 & 0.90 & 0.58\\
		\hline
		Radiomics + GRUs  & \textbf{0.96} & \textbf{0.92} & \textbf{0.95} & \textbf{0.94} & 0.82 & \textbf{0.96} & 0.75 &\textbf{0.74}\\
		\hline
	\end{tabular}
\end{table}
The classification results of the stenosis grade classification for the proposed methods and the 3D-CNN approach proposed in~\cite{zreik18b} are shown in Table~\ref{stenosisResults}.
Compared to the results reported in~\cite{zreik18b}, the performance of the 3D-RCNN approach on our dataset is lower. The main reason for this is likely the size of the respective data set, which was much smaller in our case. The proposed 2D-RCNN and radiomics approach achieved results on par with the 3D-RCNN. However, our combined approach outperformed all three other methods by a large margin (AUC 0.96 vs. 0.89 for 3D-RCNN/Radiomics+XGBoost and 0.86 for 2D-RCNN).

\subsection{Classification of Revascularization Decision}

\begin{table}[b]
	\caption{Evaluation results for revascularization decision prediction on lesion-level.}\label{revascResults}
	\begin{minipage}{\textwidth}
		\centering
		
	\begin{tabular}{ l | l | l | l | l | l | l | l | l }
		\hline
		Model/metric &  AUC & Acc & F1 & PPV & NPV & Sens & Spec & MCC\\
		\hline
		Simulated FFR best~\cite{ctffr1}\footnote[1]{Simulated FFR is compared to abnormal invasive FFR instead of revascularization decision} & 0.93& 0.86 & -- & 0.61 & 0.95 & 0.84 & 0.86 & -- \\
		\hline
		Simulated FFR worst~\cite{ctffr2}$^a$ & 0.79& 0.69 & -- & 0.56 & 0.84 & 0.61 & 0.89 & -- \\
		\hline
		\hline
		3D-RCNN \cite{zreik18b} (our data) & 0.80 & 0.76 & 0.55 & 0.45 & \textbf{0.91} & 0.72 & 0.77 & 0.42 \\
		\hline
		2D-RCNN + polar transform& 0.84 & 0.82 & 0.57 & 0.60 & 0.88 & 0.54 & 0.91 & 0.46\\
		\hline
		Radiomics + XGBoost  & 0.86 & 0.86 & 0.62 & 0.69 & 0.89 & 0.56 & \textbf{0.94} & 0.54\\
		\hline
		Radiomics + GRUs  & \textbf{0.88} & \textbf{0.87} & \textbf{0.92} & \textbf{0.90} & 0.74 & \textbf{0.95} & 0.61 &\textbf{0.60}\\
		\hline
	\end{tabular}
\end{minipage}
\end{table}
The metrics for the revascularization prediction can be seen in Table~\ref{revascResults}. 
Since there exists a lot of variance with respect to the reference standard simulated FFR, we compare our approaches to the best~\cite{ctffr1} and worst~\cite{ctffr2} results reported in the review paper of~\cite{tesche17}. Note that simulated FFR is compared to an abnormally high invasive FFR value rather than the revascularization decision in the referenced publications, with both targets being highly correlated.
The experiments in \cite{ctffr2,ctffr1} were performed on different non-publicly available data sets.
Comparing the two RCNN networks, our proposed method performed slightly better. 
This indicates that features other than the stenosis degree are relevant for the revascularization decision, and that transforming the image data into the polar space was beneficial. 
The radiomics approach outperformed both deep learning methods, while our combined approach again performed best.

\section{Discussion and Conclusion}
Identifying functionally significant stenosis in a non-invasive setup is an important task to improve clinical outcomes. We presented and compared three machine-learning methods for the prediction of stenosis degree and clinical revascularization decision based on CCTA scans: Radiomics combined with boosting trees, a convolutional recurrent neural network, and an approach that combines shape-based radiomics and recurrent neural networks. 
We were able to show that all methods were able to differentiate stenosis grade $>50$\,\% and $<50$\,\%, and reliably identify plaque lesion that were later revascularized. Across both tasks, the combined approach performed best, also compared to results reported in literature. The combined approach comes at a cost of a higher computation time of up to 2 seconds compared to only milliseconds for the RCNN approaches and requires a prior segmentation of the vessel lumen in the region of the plague segment. Still, the additional computation time does not pose a clinical limitation and the lumen segmentation is easily obtainable in an automated fashion. In contrast to this, simulated FFR requires an exact segmentation of the whole coronary tree and computation times of several minutes. For classification of revascularization, we showed that the performance of the proposed methods lies well within the range of prediction performance obtained by FFR simulation in literature. Given data with appropriate annotations, we believe that our methods would also perform well in identifying so-called culprit lesions that cause adverse cardiac events.
Interestingly, the performance difference between the combined approach and the RCNN methods leads to the conclusion that extracting the shape-based features is highly relevant for differentiating lesions, but is harder to achieve for a completely data driven CNN-based feature extractor and may require a larger training data set. If only limited data is available, the combined approach proposed here seems to be promising, as predefined features and data-driven learning are fused.
A limitation of the current study is that no simulated FFR values for the data set under investigation were available, which will be subject of future work. Additionally, the results will be validated on additional data collections that also include the invasive FFR measurements for comparison.

\subsubsection*{Disclaimer}
The methods and information here are based on research and are not commercially available.
\bibliographystyle{splncs04}
\bibliography{paper1354}
%
%
%
%
%
\end{document}